\documentstyle[preprint,aps,epsfig]{revtex}
\begin{document}
\tightenlines
\preprint{SNUTP 97-145}
\draft
\title{Electromagnetic fields 
in a 3D cavity and in a waveguide with oscillating walls
}
\author{Jeong-Young Ji\cite{e:ji}, 
Kwang-Sup Soh\cite{e:soh}
}
\address{Department of Physics Education, Seoul National University, 
Seoul 151-742, Korea
}
\author{Rong-Gen Cai\cite{e:cai}
}
\address{Center for Theoretical Physics,
Seoul National University, Seoul, 151-742, Korea}
\author{
Sang Pyo Kim\cite{e:kim}
}
\address{Department of Physics, 
Kunsan National University,
Kunsan 573-701, Korea}

\maketitle
\begin{abstract}
We consider classical and quantum electromagnetic fields in a 
three-dimensional (3D) cavity and in a waveguide with oscillating 
boundaries of the frequency $\Omega $. The photons created by the 
parametric resonance are distributed in the wave number space around 
$\Omega/2 $ along the axis of the oscillation. When classical waves 
propagate along the waveguide in the one direction, we observe the 
amplification of the original waves and another wave generation in the 
opposite direction by the oscillation of side walls. This can be 
understood as the classical counterpart of the photon production. In the 
case of two opposite walls oscillating with the same frequency but with a 
phase difference, the interferences are shown to occur due to the phase 
difference in the photon numbers and in the intensity of the generated 
waves.
\end{abstract}


\narrowtext

The Casimir effect~\cite{Casimir48} is a macroscopic manifestation of the 
change in the zero-point electromagnetic energy due to the walls. The 
time-varying boundary conditions induce the change of the vacuum states 
for the quantum electromagnetic fields and the difference between initial 
and final vacuum states results in the photon production. This dynamical 
Casimir effect provides the possibility to observe experimentally the 
vacuum change of quantum fields. This phenomenon has been extensively 
studied when the one of the walls 
oscillates~\cite{DoKM90,SaSW94,KlimovA97,SchPS97}: For an almost 
sinusoidal movement of the mirror, using the formalism invented by 
Moore~\cite{Moore70} and developed 
by Fulling and Davies~\cite{FullingD76}, the quantum energy density has 
been calculated~\cite{Law94,ColeS95}. For a harmonic oscillation of the 
mirror M\'{e}plan and Gignoux have shown that a set of frequencies of the 
oscillating walls leads to an exponential growth of the energy of a 
wave~\cite{MeplanG96} and the exponential growth of the number of 
generated photons can be easily understood from the Floquet's 
theorem~\cite{JiS97}. The scattering approach is used in analyzing the 
motion induced radiation from a vibrating cavity with partly transmitting
mirror(s)~\cite{Jaekel,LambJR96}. For the small oscillation of the walls, 
the perturbation approach~\cite{J} have been developed to calculate the 
time evolution of the electromagnetic field in the instantaneous 
basis~\cite{Law95,Do95,Do96,DoK96}. 

The aim of this Letter is to examine the photon production in a 3D cavity 
and to consider the classical electromagnetic fields propagating in a 
waveguide with oscillating walls. The photons created by the parametric 
resonance are distributed in the wave number space around the half of the 
oscillation frequency along the axis of the oscillating motion. We shall 
show that if we transmit the classical waves into the waveguide with 
oscillating walls, the waves are amplified and there are generated waves 
propagating in the opposite direction, which corresponds to the photon 
production in the quantum theory. When two walls oscillate we find the 
interference phenomena in the photon numbers and in the intensity of the 
generated waves. 

Assuming that the electric field ${\bf E} ( {\bf r}, t ) $ is polarized 
in the $z$ direction,
we may write~\cite{Moore70}
\begin{eqnarray}
{\bf A} &=& A( x, y, t ) {\bf \hat{z}},
\nonumber \\
{\bf E} &=& E {\bf \hat{z}} =
- \frac{{\partial A}}{\partial t} {\bf \hat{z}},
\\
{\bf B} &=& \frac{{\partial A}}{\partial y} {\bf \hat{x}} 
- \frac{{\partial A}}{\partial x} {\bf \hat{y}} . 
\nonumber
\end{eqnarray}
Consider a rectangular cavity with sides $q_{x} (t), L_{y} $ and $L_{z} , $ 
where the one of the wall oscillates for a time interval $0 < t < T $ with 
a small amplitude $( \epsilon \ll  1) $ according to
\begin{equation}
q_{x} (t) = L_{x} ( 1 + \epsilon \sin \Omega t ) .
\end{equation}
In this cavity the field operator can be expanded 
\begin{equation}
A = \sum_{\bf n} [ 
b_{\bf n} \psi_{\bf n} +
b_{\bf n}^\dagger \psi_{\bf n}^* ]
\end{equation}
using the following instantaneous basis
\begin{equation}
\psi_{\bf n} (x,y| q_{x} (t) ) = \sum_{\bf k}
Q_{\bf nk} \varphi_{\bf k} (x,y,t)
\end{equation}
where
\begin{equation}
\varphi_{\bf k} (x,y | q_{x} (t) )
= \frac{2}{\sqrt{q_{x} L_{y} L_{z}}} \sin \frac{{\pi k_{x} x}}{q_{x}}
\sin \frac{{\pi k_{y} y}}{L_{y}} ,
\end{equation}
with $k_{x} , k_{y} = 1, 2, 3, ... . $
 From the Maxwell's equations or the wave equation for $\psi_{\bf n} $ we 
have
\begin{eqnarray}
{\ddot{Q}}_{\bf nk} 
&=& 
- \omega_{\bf k}^2 Q_{\bf nk}
+ 2 \epsilon ( \pi k_{x} / L_{x} )^{2} \sin \Omega t 
Q_{\bf nk}
\nonumber \\
&& + 
2 \epsilon \Omega \cos \Omega t 
\sum_{\bf j} g_{\bf kj} {\dot{Q}}_{\bf nj}
- \epsilon \Omega^{2} \sin \Omega t 
\sum_{\bf j} g_{\bf kj} Q_{\bf nj} 
\nonumber \\
&& + O( \epsilon^{2} )
\label{EOM}
\end{eqnarray}
with
\begin{equation}
g_{\bf jk} 
=(-1)^{j_{x} +k_{x}} \frac{{2j_{x} k_{x}}}{k_{x}^{2} - j_{x}^{2}} 
\delta_{j_{y} k_{y}}  ~ ( j_{x} \neq k_{x} ) ,
\label{g}
\end{equation}
where $g_{\bf jk} = 0  $ for $j_{x} = k_{x} . $
Using the perturbation method developed in Ref.~\cite{J}, the solution 
can be written as
\begin{equation}
Q_{\bf nk} = Q_{\bf nk}^{(0)} 
+ \epsilon Q_{\bf nk}^{(1)} + \cdots
\end{equation}
where
\begin{equation}
Q_{\bf nk}^{(0)} (t) = \frac{{ e^{-i \omega_{\bf k} t} }}{\sqrt{2 
 \omega_{\bf k}}}
\delta_{\bf nk} 
\label{sol0}
\end{equation}
and
\begin{equation}
Q_{\bf nk}^{(1)} (t) =
\sum_{\sigma, s=\pm}
w_{\bf k \sigma , n-}^{s} 
\frac{e^{\sigma i \omega_{\bf k} t}}{\sqrt{2 \omega_{\bf k}} } 
\int_{0}^{t} dt' e^{-i 
(\sigma \omega_{\bf k} - s \Omega + \omega_{\bf n}) t'} 
\label{sol1}
\end{equation}
with
\begin{equation}
w_{\bf k \sigma, n \sigma'}^{s} 
= \sigma
\left[
\Omega g_{\bf kn} \sqrt{ \frac{\omega_{\bf n}}{\omega_{\bf k}} } 
\left( \frac{{s \Omega}}{4 \omega_{\bf n} } + \frac{{ \sigma' }}{2} \right)
- s \frac{{ (k_{x} \pi/ L_{x} )^{2} }}{2 \omega_{\bf k} } \delta_{\bf kn}
\right] .
\label{vpm}
\end{equation}
Here we note that the zeroth order solution describes the field operator 
at the static situation: 
\begin{equation}
A = \sum_{\bf n} [ 
b_{\bf n} \phi_{\bf n} +
b_{\bf n}^\dagger \phi_{\bf n}^* ]
\end{equation} 
where $\phi_{\bf n} = \frac{1}{ \sqrt{2 \omega_{\bf n} } } 
e^{-i \omega_{\bf n} t} \varphi( x, y | L_{x} ) $. After some interval 
$T$ of the oscillation of the wall, the Heisenberg field operator can be 
written as
\begin{equation}
A = \sum_{\bf n} [ 
a_{\bf n} \phi_{\bf n} +
a_{\bf n}^\dagger \phi_{\bf n}^* ]
\end{equation}
where
\begin{equation}
a_{\bf k} = \sum_{\bf n} 
[ b_{\bf n} \alpha_{\bf nk} +
b_{\bf n}^\dagger \beta_{\bf nk}^* ] ,
\end{equation}
with
\begin{equation}
\sum_{\bf n} 
( | \alpha_{\bf nk} |^{2} - 
| \beta_{\bf nk} |^{2} ) = 1 .
\label{ab1}
\end{equation}
The Bogoliubov coefficient $\beta_{\bf nk} $ can be read from the 
solution $Q_{\bf n k}^{(1)} $  to the leading order in $\epsilon $ by 
retaining dominant terms only $( \omega T \gg  1 ) $:
\begin{equation}
\beta_{\bf nk} = \epsilon T w_{\bf k+ , n-}^+
\delta_{ \omega_{\bf n}, \Omega - \omega_{\bf k}} .
\label{bnk}
\end{equation}
Hereafter we introduce the bar notation for the wave number vector: 
$\bar{\bf n} $ denotes the wave number vector corresponding to $\bf n $ 
or 
in the components, 
$( \bar{n}_{x} , \bar{n}_{y} ) = ( n_{x} \pi / L_{x} , n_{y} \pi / L_{y} ) 
 . $
Noting that the resonance conditions 
\begin{equation}
\omega_{\bf n} = \Omega - \omega_{\bf k} 
~ {\rm and} ~ \bar{n}_{y} = \bar{k}_{y} 
\label{rc}
\end{equation}
can be explicitly written as~\cite{nx:int} 
\begin{equation}
\bar{n}_{x}^{2}  = 
\bar{k}_{x}^{2} -2 \Omega \omega_{\bf k} + \Omega^{2} ,
\label{nx:kx}
\end{equation} 
we have the following number distribution in the created photons:
\begin{eqnarray}
N_{\bf k} &=& \sum_{\bf n} |\beta_{\bf n k}|^{2} 
\nonumber \\
&=& 
\left( \frac{{\epsilon T}}{2} \right)^{2} 
\frac{{  \bar{k}_{x}^{2} [ \bar{k}_{x}^{2} -2 \Omega \omega_{\bf k}+ 
 \Omega^{2} ]}}{ \omega_{\bf k} (\Omega-\omega_{\bf k})} 
\label{Nk}
\end{eqnarray}
This distribution of created photons is anisotropic in the wave number 
space (see Fig. 1) and the number of created photons is maximal at the 
nearest neighbor of
\begin{equation}
\bar{k}_{x} = \frac{\Omega}{2}  ~ {\rm and} ~ \bar{k}_{y} = 0,
\label{maxi}
\end{equation}
which come from the conditions 
${\partial N_{\bf k}} / {\partial \bar{k}_{x}} = 0 $ and 
${\partial N_{\bf k}} / {\partial \bar{k}_{y}} = 0 $. In fact, the second 
condition in (\ref{maxi}) means that $k_{y} = 1 $ because $k_{y} = 0 $ 
means the vanishing field. Note that the maximum number of photons are 
created for $\omega_{\rm m a x} = \Omega / 2 $ which is the characteristic 
condition of the parametric resonance near on the axis of oscillation 
(with the minimum $\bar{k}_{y} .) $

Now we extend the results to the case when the left and the right walls 
oscillate with the frequencies $\Omega_{L} $ and $\Omega_{R} $ 
respectively:
\begin{equation}
N_{\bf k} =
N_{\bf k}^L + N_{\bf k}^R - 
(-1)^{k_{x} + n_{x}} 2 \sqrt{ N_{\bf k}^L } \sqrt{ N_{\bf k}^R } \cos 
 \phi 
\delta_{\Omega_{L} , \Omega_{R}} .
\label{Nk:LR}
\end{equation}
where $N_{\bf k}^{L} $ and $N_{\bf k}^{R} $ are obtained by replacing 
$\Omega $ in (\ref{Nk}) with $\Omega_{L} $ and $\Omega_{R} $, 
respectively. Here $\phi $ is the initial phase difference between two 
oscillations of the walls and $n_{x}  $ is a positive integer satisfying 
(\ref{nx:kx})~\cite{nx:int}. When $\Omega_{L} \neq \Omega_{R} , $ the 
number of generated photons by the parametric resonance is the sum of the 
photon numbers generated when the left and the right wall oscillates 
separately. When $\Omega=\Omega_{L} = \Omega_{R} , $ there is an 
additional interference term depending on the mode number of $x$ component 
and the phase difference $\phi $. This is just the interference phenomenon 
found in the 1D case~\cite{JJS97}. It is worth noting that when $\phi = 0 $ 
or $\phi = \pi $ whether the interference is constructive or destructive 
is determined by $x$-component mode numbers only. Unlike the 1D case, 
$n_{x} + k_{x} $ does not represent the ratio of the oscillation 
frequency $\Omega $ to the fundamental mode frequency 
$\omega_{(1,1)} = \sqrt{( \pi/L_{x} )^{2} + ( \pi/L_{y} )^{2}} . $ But 
for $L_{y} \gg  L_{x}  $ and $\bar{k}_{y} \approx 0 $, $n_{x} + k_{x} $ is 
an integer close to $\gamma = \Omega / \omega_{(1,1)} . $ In this case we 
have a constructive (destructive) interference when $\phi = \pi $ for 
$\gamma = 2, 4, ... $$\gamma = 1, 3, .... $) and when $\phi = 0 $ for 
$\gamma = 1, 3, .... $$\gamma = 2, 4, ... $). For a general 3D mode, it 
is possible to have a constructive (destructive) interference when 
$\phi = \pi $ for an odd (even) $\gamma $ since the mode frequency 
depends on not only the $x$-mode number but also the $y$-mode number.

We now turn to the classical electromagnetic waves propagating in the 
positive $y$-direction in the rectangular cavity with an oscillating wall. 
When the wall of a cavity is static the right-going wave is described by 
\begin{equation}
A(x,y,t) = 
\sum_{\bf n}
f_{\bf n} 
{\cal N}_{\bf n}
\cos ( \bar{n}_{y} y - \omega_{\bf n} t ) 
\sin \bar{n}_{x} x
\label{A0:go}
\end{equation}
with the normalization constant 
${\cal N}_{\bf n} = \sqrt{ 2 / {\omega_{\bf n} \pi L_{x}  L_{z} }}  . $
Here the sum over ${\bf n} $ denotes the sum over 
$\bar{n}_{x} = n_{x} \pi / L_{x}
~(n_{x} = 1,2,3,...) $ and the integration over $\bar{n}_{y} $,  and 
$f_{\bf n} $ is a real distribution function of the wave number vector 
$\bar{n} = \bar{n}_{x} {\bf \hat{x}} + \bar{n}_{y} {\bf \hat{y}}. $  
There is no boundary along the $y$-direction and $\bar{n}_{y} $ can be 
regarded as a continuum limit of $n_{y} \pi / L_{y} $ 
$(L_{y} \rightarrow \infty ). $ Introducing the propagating instantaneous 
basis
\begin{equation}
\varphi_{\bf k} (x,y| q_{x} (t) )
= \frac{1}{\sqrt{ \pi q_{x} (t) L_{z} }}  
\sin \frac{{\pi k_{x} x}}{q_{x} (t)}
e^{ i \bar{k}_{y} y} ,
\label{phk:go}
\end{equation}
the classical vector potential at any time can be expanded as 
\begin{equation}
A(x,y,t) = 
\sum_{\bf n} [ f_{\bf n}
\sum_{\bf k} 
( Q_{\bf n k} \varphi_{\bf k}
+ {\rm H.c.} ) ] .
\label{A:go}
\end{equation}
With the initial condition (\ref{sol0}) and the static-wall solution with 
$q_{x} (t) = L_{x} $ in (\ref{phk:go}), the field (\ref{A:go}) becomes 
(\ref{A0:go}). The time evolution of $Q_{\bf nk} (t) $ is given by solving 
the same equation  (\ref{EOM}) and we have the same solution. Then for the 
oscillating wall we have the following wave :
\begin{eqnarray}
A(x,y,t) &=& 
\sum_{\bf n, k}
f_{\bf n} 
\alpha_{\bf nk} {\cal N}_{\bf k}
\cos ( \bar{k}_{y} y - \omega_{\bf k} t ) 
\sin \bar{k}_{x} x
\nonumber \\
&&
+ 
\sum_{\bf n, k}
f_{\bf n} 
\beta_{\bf nk}
{\cal N}_{\bf k}
\cos ( \bar{k}_{y} y + \omega_{\bf k} t ) 
\sin \bar{k}_{x} x ] .
\label{AT:go}
\end{eqnarray}
where $\beta_{\bf nk} $ is given by (\ref{bnk}) to the leading order in 
$\epsilon $.
Thus we have the left-going wave in the $y$-direction induced by the 
$x$-directional oscillation of the wall, with the amplitude  of ${\bf k} $
th mode being proportional to $\sum_{\bf n} f_{\bf n} \beta_{\bf nk}
 $ 
with $\left| \sum_{\bf n} f (\bar{\bf n} ) \beta_{\bf nk} \right|^{2}
= f (\sqrt{ \bar{k}_{x}^{2} -2 \Omega \omega_{\bf k} + \Omega^{2} }, 
\bar{k}_{y} ) 
N_{\bf k} , $
where $f( \bar{\bf n} ) $ denotes $f_{\bf n}  $ and $N_{\bf k} $ is given 
by Eq.~(\ref{Nk}). When the two side-walls oscillate, we find the 
interference phenomena again as in the quantum field case.

Before proceeding on analyzing our results, let us survey other works for 
the 3D cavity with oscillating walls. It has been suggested that a 
fantastic amount of photons can be generated in a 3D cavity with one plate 
being performed periodic instantaneous jumps between two stationary 
positions~\cite{SaSW94}. However, this result was obtained by neglecting 
the terms coupled to other frequency modes in Eq.~(\ref{EOM}), as pointed 
out in Ref.~\cite{DoK96}, and the cases not satisfying Eq.~(35) in 
Ref.~\cite{SaSW94} which correspond to the condition of parametric 
resonance~\cite{JK95} was neglected. In fact, such large number can be 
obtained from the ultraviolet divergence for the instantaneous jump of the 
frequency. In Ref.~\cite{Do95,DoK96}, the 3D problem has been reduced to a 
decoupled single parametric oscillator:  Using the ansatz 
$Q_{\bf k} = \xi_{\bf k} (\epsilon t) e^{-i \omega_{\bf k}t}
+ \eta_{\bf k} (\epsilon t) e^{i \omega_{\bf k}t} $ together with the 
assumption that $\xi_{\bf k} $ and $\eta_{\bf k} $ are slowly varying 
functions of time, and  averaging over fast oscillations, the coupling 
terms [second line in Eq. (\ref{EOM})] have been neglected again in the 
effective theory. In our case, the coupling terms are also considered and 
they affect the calculation of the Bogoliubov coefficient $\beta_{\bf nk} . $ 

One may wonder if the resonance condition (\ref{nx:kx}) cannot be 
satisfied due to the discreteness of the frequency in the cavity. As 
stated~\cite{nx:int}, small deviation from the resonance condition is 
admitted and it can be satisfied in the continuum limit 
$(L_{x} \ll  L_{y} ) $. Thus, the propagating wave in the wave guide 
$( L_{y} \rightarrow \infty ) $ provides a good experimental situation to 
observe the generation of the wave or the photon production. The intensity 
of the generated wave is the order of the intensity of the incident wave 
multiplied by the number of produced photons in the quantum theory. After 
a very short time, we may observe an amount of created left-going wave if 
we prepare the incident right-going wave satisfying the resonance 
condition (\ref{rc}). For an experimental situation to observe $N$ photons 
per second ($N=10$ in the Ref.~\cite{LambJR96}, $N=600$ in the 
Ref.~\cite{DoK96}), it takes only $1/N$ second to get the same intensity 
of the generated wave as the incident wave, then the incident right-going 
wave will be also amplified by the order of $1+N$ according to 
(\ref{ab1}). This phenomenon can be regarded as the classical counterpart 
of the photon production in the quantum theory and it is easily observable 
in the experimental situation.

This work was supported by the Center for Theoretical Physics (S.N.U.), 
and the Basic Science Research Institute Program, Ministry of Education 
Project No. BSRI-97-2418. One of the authors (J.Y.J.) would like to thank 
Dr. J. H. Cho for very helpful discussions.

\begin{figure}[htb]
\centerline{\epsfig{figure=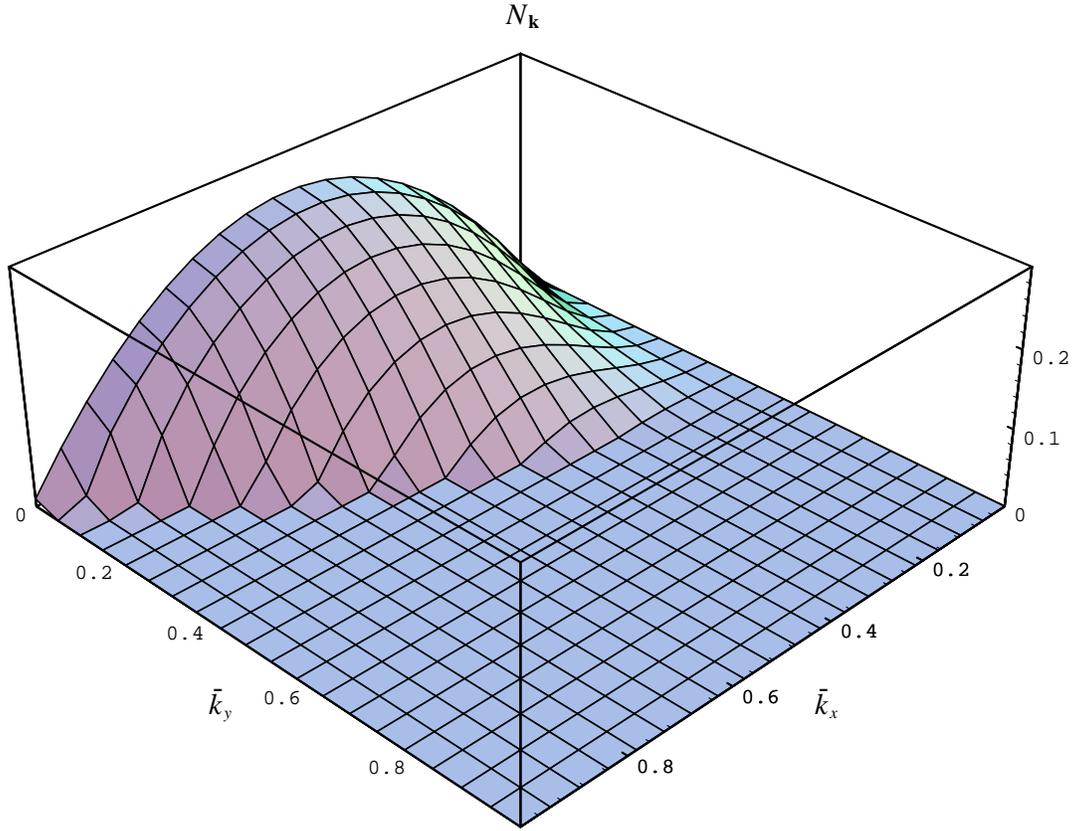,width=\hsize,angle=0}}
\caption{
The distribution of created photons in the rectangular cavity 
when the side at $ x=q_x (t) $ oscillates sinusoidally with a small 
amplitude. $N_{\bf k} $ is denoted in units of $(\epsilon \Omega T/2)^{2} , $ 
and $\bar{k}_{x} $ and $\bar{k}_{y} $ are denoted in units of $\Omega . $
}
\label{fig1}
\end{figure}


\begin{references}
\bibitem[*]{e:ji} Electronic address: jyji@phyb.snu.ac.kr, 
\bibitem[\dag]{e:soh} Electronic address: kssoh@phya.snu.ac.kr
\bibitem[\ddag]{e:cai} Electronic address: cairg@ctp.snu.ac.kr
\bibitem[\S]{e:kim} Electronic address: sangkim@knusun1.kunsan.ac.kr

\bibitem{Casimir48} H. B. G. Casimir, 
Proc. K. Ned. Akad. Wet. {\bf 51}, 793 (1948).

\bibitem{DoKM90} V. V. Dodonov, A. B. Klimov, 
and V. I. Man'ko, Phys. Lett. A {\bf 149}, 225 (1990). 
\bibitem{SaSW94} E. Sassaroli, Y. N. Srivastava, and A. Widom, Phys. Rev. 
A {\bf 50}, 1027 (1994).
\bibitem{KlimovA97} A. B. Klimov, and V. Altuzar, 
Phys. Lett. A {\bf 226}, 41 (1997).
\bibitem{SchPS97} R. Sch\"{u}tzhold, G. Plunien, and G. Soff,
quant-ph/9709008.

\bibitem{Moore70} G. T. Moore, 
J. Math. Phys. (N.Y.) {\bf 11}, 2679 (1970).
\bibitem{FullingD76} S. Fulling and P. Davies, Proc. R. Soc. London
A {\bf 348}, 393 (1976).
\bibitem{Law94} C. K. Law, Phys. Rev. Lett. {\bf 73}, 1931 (1994).
\bibitem{ColeS95} C. K. Cole, W. C. Schieve, 
Phys. Rev. A {\bf 52}, 4405 (1995).

\bibitem{MeplanG96} O. M\'{e}plan, and C. Gignoux, 
Phys. Rev. Lett. {\bf 76}, 408 (1996).
\bibitem{JiS97} J. Y. Ji and K. S. Soh, 
in {\it Proceedings of the 5th Korean-Italian Symposium 
on Relativistic Astrophysics}
(to be published in J. Kor. Phys. Soc.).

\bibitem{Jaekel} M. T. Jaekel and S. Reynaud, 
Quantum Opt. {\bf 4}, 39 (1992); 
J. Phys I (France) {\bf 2}, 149 (1992).
\bibitem{LambJR96} A. Lambrecht, M.-T. Jaekel, 
and S. Reynaud, Phys. Rev. Lett. {\bf 77}, 615 (1996).

\bibitem{J} J. Y. Ji, H. H. Jung, J. W. Park, and K. S. Soh,
quant-ph/9706007 (to be published in Phys. Rev. A).

\bibitem{Law95} C. K. Law, Phys. Rev. A {\bf 51}, 2537 (1995). 
\bibitem{Do95} V. V. Dodonov, 
Phys. Lett. A {\bf 207}, 126 (1995).
\bibitem{Do96} V. V. Dodonov, 
Phys. Lett. A {\bf 213}, 219 (1996).
\bibitem{DoK96} V. V. Dodonov, A. B. Klimov, 
Phys. Rev. A {\bf 53}, 2664 (1996). 

\bibitem{nx:int} There may not exist an integer $n_x$ satisfying 
(\ref{nx:kx}) due to the discreteness of the frequency and in this case
there are no photons created by the parametric resonance. 
But for the case $L_{x} \ll  L_{y}  $, the frequency can be regarded as a 
continuum
and in this case the resonance condition will be fulfilled 
by an integer $n_x$. Further the condition of parametric resonance
admits some discrepancy as seen from 
the solutions of the Mathieu equation.
\bibitem{JJS97} J. Y. Ji, H. H. Jung, and K. S. Soh,
quant-ph/9709046.
\bibitem{JK95} J. Y. Ji and J. K. Kim, Phys. Lett. A {\bf 208}, 25 
(1995).
\end{references}
\end{document}